# MESURE SIMULTANEE DES REPONSES IMPULSIONNELLES EN MACRODIVERSITE

**Karim ZAYANA, Daniel DUPONTEIL**

France Télécom - Centre National d'Etudes des Télécommunications
38-40 rue du Général Leclerc 92794 Issy les Moulineaux Cedex 9

RESUME

La macrodiversité d'émission dans les systèmes radiomobiles est une technique dans laquelle le terminal est en liaison parallèle avec deux stations de base. Elle est un élément fondamental du fonctionnement des réseaux mobiles de troisième génération utilisant l'AMRC. (IS 95, UMTS, etc.). Une représentation des phénomènes de propagation mis en jeu nécessite une modélisation conjointe des deux canaux de transmission. L'article présente une méthode très simple de mesure simultanée de leurs réponses impulsionnelles, mise au point dans ce but. Cette méthode a fait l'objet d'un brevet, /1/. Elle peut plus généralement s'appliquer à un nombre quelconque de canaux et devrait permettre la réutilisation des sondeurs monocanaux déjà existants.

## 1. Introduction

La macrodiversité d'émission dans les systèmes radiomobiles est une technique dans laquelle le terminal est en liaison parallèle avec deux stations de base.

Un nombre croissant de systèmes de transmission à hauts débits, tels l'IS95 ou l'UMTS à venir utilisent la macrodiversité pour lutter contre les affaiblissements rapides et les effets de masque, et adoucir les transitions entre cellules. Il est donc indispensable de réaliser des campagnes de mesure large bande en macrodiversité afin de concevoir et déployer au mieux les nouveaux réseaux.

## 2. Mesures en Macrodiversité

### 2.1. Le sondage de canal

Les sondeurs de canaux traditionnels fonctionnent par compression d'impulsion. Une séquence de durée $T$ et présentant de bonnes propriétés d'autocorrélation est synthétisée numériquement en bande de base où elle occupe une bande de largeur $B$.

Cette séquence est par exemple une $m$-séquence, constituée d'un train pseudo-aléatoire de 0 et de 1. Dans des versions plus élaborées, ce peut aussi être une séquence à valeurs complexes optimisée de manière à ce que son enveloppe après filtrage reste encore quasiment constante, /2/ un séquence maximale. Dans ces conditions, et pour éviter tout problème entre voies I et Q lors de la modulation, le passage de la bande de base vers la fréquence intermédiaire s'opère numériquement, figure 1.

ABSTRACT

Macroscopic Diversity uses signals from two -sometimes more-nearby ports in a way to mitigate the effects of shadow and fast fading. It is an essential component of emerging spread spectrum systems (IS95, UMTS, etc.). To provide a good representation of the propagation phenomena involved in macrodiversity, one need to characterize jointly both transmission channels. To this end, the article presents a clever method to achieve hign quality and simultaneous mesasurements of two impulse response channels. The method proposed is new and has been patented, /1/. More generally, it can be applied to an any number of channels, and should allow to reuse already existing single-channel sounders.

Le signal est ensuite porté à la fréquence d'émission $f_0$. Par modulation et filtrages, figure 1.

En réception, le signal est ramené en bande de base, puis corrélé avec la séquence d'origine, ce qui fournit une estimée de la réponse du canal. Le traitement peut tout aussi bien s'opérer fréquentiellement, par inversion de Wiener, /3/.

En pratique, on enregistre d'abord un fichier câble de manière à éliminer des mesures le filtrage dû aux appareils de transmission, /3/.

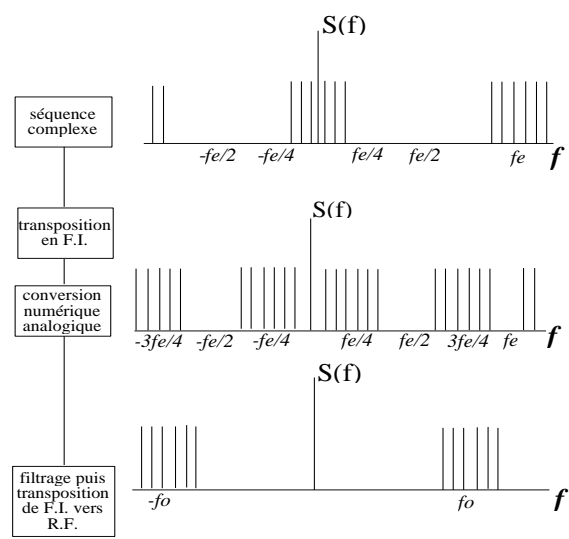

**Figure 1 : chaîne d'émission d'un sondeur large bande**

## 2.2. Les campagnes en Macrodiversité

Les campagnes de mesures en macrodiversité sont jusqu'à présent peu répandues car de mise en œuvre délicate. On les classe en deux catégories :

*Les campagnes « temps maître »*. Le mobile émet un signal sondeur qui est reçu par les deux stations. Deux récepteurs sont donc nécessaires. Une personne est présente sur chaque site pour enclencher le début puis la fin des mesures. Les acquisitions s'effectuent à un rythme régulier :elles sont commandées par une horloge en réception. La méthode est dite « temps maître ».

Dans cette configuration, le véhicule doit rouler à vitesse constante afin de reconstituer l'emplacement exact des points de mesure. Cela n'est possible que sur des petits tronçons, et certains parcours peuvent être exclus à cause de la circulation routière. S'y ajoutent des difficultés de synchronisation puisque les instants de déclenchement et d'arrêt des enregistrements doivent être communs aux deux sites.

*Les campagnes «distance maître »*. L'émission a lieu depuis chaque site, la réception s'effectuant au mobile. Un capteur fixé à la roue du mobile contrôle les acquisitions, ce qui rend aisée la localisation des points de mesure. La configuration est dite « distance maître ».

Une première solution, conceptuellement la plus simple, consiste à émettre le même signal sur les deux liaisons. Cela a pour effet de superposer des réponses impulsionnelles des deux voies en réception. Mais il est en pratique très difficile de garantir leur non chevauchement, /4/.

Pour y remédier, G.Kadel a proposé d'introduire un Doppler artificiel d'une vingtaine de Hertz sur la seconde voie, /5/. Le signal composite reçu reste démodulé à la fréquence porteuse du premier émetteur. Après traitement, les deux réponses impulsionnelles se trouvent encore superposées. Si le mobile reste stationné, l'observation d'un grand nombre de mesures rapprochées en temps fait apparaître une fluctuation de la seconde réponse alors que la première ne bouge pas. Cela permet de séparer les deux voies après filtrage.

Il faut donc enregistrer un nombre suffisant de mesures consécutives en chaque point, ce qui demande beaucoup d'espace de stockage et contraint le mobile à s'arrêter ou à rouler très doucement. Le traitement a postériori est ensuite gourmand en calculs pour les deux voies.

## 3. Une nouvelle méthode de mesures

### 3.1. Descriprion de la méthode

Nous proposons une nouvelle solution « distance maître » pour mesurer simultanément les réponses impulsionnelles de plusieurs canaux radiomobiles.

La méthode s'adapte très bien aux sondeurs déjà existants. Elle permet un traitement en temps réel, puisque chaque mesure à un instant $t$ permet de calculer les réponses impulsionnelles des deux canaux de ce même instant. Il n'y a donc plus de contrainte sur la vitesse du véhicule. Les calculs sont d'une complexité double de ceux effectués sans diversité.

La méthode s'étend à plus de deux canaux. Elle est réutilisable en microdiversité d'espace ou de polarisation.

L'idée est d'émettre depuis les stations deux signaux qui, en bande de base, sont périodiques, orthogonaux et couvrent simultanément la même bande à sonder. Cela est par exemple réalisé lorsque les raies spectrales des deux signaux émis sont entrelacées, figure 2.

Pour y parvenir, il suffit d'utiliser les mêmes séquences sondeur de durée $T$ aux deux émetteurs, puis de décaler fréquentiellement de $1/2T$ la porteuse du second émetteur par rapport à celle du premier, figure 3. On peut également, intégrer ce décalage fréquentiel directement lors de la synthèse numérique de la séquence émise en voie 2. Auquel cas, les porteuses des deux émetteurs sont aux mêmes fréquences.

La période du signal bande de base émis en voie 2 s'en trouve toujours doublée. L'orthogonalité entre les deux liaisons est satisfaite sur toute durée $2T$.

En réception et après démodulation, on corrèle circulairement l'enregistrement successivement avec un motif de durée $2T$, extrait du signal émis en voie 1, puis avec un motif de même durée $2T$ extrait du signal émis en voie 2. La première opération fournit à quelques détails près la réponse impulsionnelle du premier canal tandis que la seconde corrélation donne la seconde réponse. On peut tout aussi bien imaginer un traitement par inversion de Wiener en raisonnant en fréquence et en sélectionnant une raie sur deux selon la voie à étudier.

Détaillons les calculs dans ce deuxième cas, la méthode par inversion étant la plus simple à exposer. Les raisonnements sont présentés en bande de base, c'est à dire que la fréquence de la porteuse du premier émetteur est choisie comme référence.

Notons $s_1$ et $s_2$ les signaux sondeurs émis depuis les sources 1 et 2 et $S_1$ et $S_2$ leurs transformées de Fourier, figure 2. Le signal $s_1$ est $T$-périodique. Son spectre est constitué de raies aux fréquences $k/T$. Les raies porteuses d'information sont réduites au nombre $2N$ par exemple après filtrage à l'émission, ainsi $k \in [-N\,;N-1]$. On pose alors $N/T=B/2$ et $S_1$ occupe donc la bande $[-B/2\,;B/2[$ de largeur $B$. Le second émetteur est décalé en fréquences de $1/2T$ par rapport au premier. Le signal $s_2$ est $2T$-périodique. Son spectre est constitué de raies consécutives distantes de $1/T$ et de la forme $k/T+1/2T$ avec $k \in [-N\,;N-1]$. Il occupe aussi la bande $[-B/2\,;B/2[$.

Le signal composite reçu est observé sur une durée $2T$. L'échantillonnage périodise son spectre. On ne constate pas de repliement lorsque la fréquence d'échantillonnage $f_e$ est suffisante, /6/. L'observation sur la durée $2T$ a pour effet de convoluer la transformée de Fourier du signal reçu par un sinus cardinal au lobe principal de largeur $1/T$. Deux raies distinctes appartenant au spectre fondamental n'interfèrent pas aux fréquences $k/2T$, résultat de la distribution des zéros du sinus cardinal. Une raie harmonique n'interfère pas non plus avec les raies du spectre fondamental à ces mêmes fréquences dès que $f_e$ est supérieure à $B$ et multiple de $1/2T$, par exemple $f_e = 2N/T = B$ ou encore $f_e = 2B$ lorsqu'on souhaite réaliser numériquement le retour de la F.I. à la bande de base. Choisissons dans la suite $f_e=B$.

Sur la durée $2T$ d'observation, on récupère donc $f_e.2T=2BT=4N$ échantillons. Une TFD (Transformée de Fourier Discrète) de ces échantillons calcule aux fréquences $k/2T$ la transformée de Fourier $R_{e,2T}$ du signal reçu $r$ échantillonné à la fréquence $f_e$ et analysé sur une durée $2T$. Pour les raisons évoquées plus haut, les valeurs renvoyées en ces points par la TFD correspondent précisément aux amplitudes des raies du spectre $R$ du signal reçu $r$.

La suppression des raies impaires et le calcul de transformée inverse $TFD^{-1}\left(R\left(\frac{k}{T}\right)/S_1\left(\frac{k}{T}\right)\right)_{k\in[-N;N-1]}$ donne une observation sur la durée $T$ de la réponse impulsionnelle du canal 1 restreint à la bande $B$, $T$-périodisée et échantillonnée à la fréquence $f_e$ (calculs en Annexe). Si la durée $T$ excède la durée d'étalement des retards, on en déduit une bonne estimation de la première réponse.

De même, la suppression des raies paires et le calcul de $TFD^{-1}\left(R\left(\frac{k}{T}+\frac{1}{2T}\right)/S_2\left(\frac{k}{T}+\frac{1}{2T}\right)\right)_{k\in[-N;N-1]}$ donne une estimée de la réponse du canal 2 affectée d'un déphasage lentement variable (voir Annexe), qu'on élimine après coup.

Remarquons enfin que les TFD peuvent être avantageusement remplacés par des FFT (Fast Fourier Transform) grâce à des astuces de calcul ramenant les inversions à des convolutions. Cela rend l'algorithme proposé plus rapide et donc plus attractif.

La méthode présentée pour deux émetteurs se généralise à $p$ émetteurs ($p \geq 2$) en inversion comme en corrélation. On entrelace en fréquence les sources aux multiples de la quantité $1/pT$. Le rythme d'échantillonnage en réception peut être conservé, soit $f_e = 2N/T = B$. Le temps d'acquisition vaut $pT$. On estime le $n$-ième canal ($1 \leq n \leq p$) en extrayant les raies numérotées $n$ modulo $p$, en revenant en temps, puis en corrigeant le déphasage $e^{-j\pi(n-1)t/pT}$ (voir Annexe). Notons en effet

L'approche du problème est radicalement différente des solutions « distance maître » traditionnelles. Le décalage en fréquence de $1/2T$ ne peut être interprété comme un Doppler artificiel, ce dont témoigne le traitement original qui est proposé. Ce décalage vaut par exemple 25 KHz pour $T=20$ $\mu s$. Un tel ordre de grandeur n'a aucun rapport avec la vingtaine de Hertz en question auparavant. Pas plus d'ailleurs que le temps d'observation nécessaire au calcul d'une réponse impulsionnelle, qui n'est ici que de 40 $\mu s$.

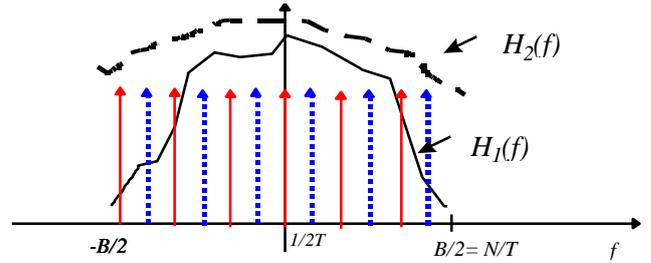

**Figure 2 : Entrelacement des raies**

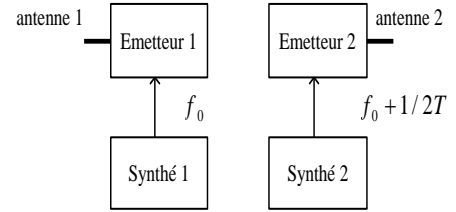

**Figure 3 : Schéma d'émission**

### 3.2. Expérimentations et résultats

La méthode proposée à été validée puis expérimentée par le CNET en situations de macrodiversité et de microdiversité.

Nous donnons ci-après un exemple de mode opératoire en macrodiversité dans la bande des 2,2 GHz. Pour l'expérimentation décrite, les deux émetteurs ont été placés au dessus des toits à une distance de 600 m l'un de l'autre dans le centre ville de Mulhouse (Haut Rhin).

La première source émet périodiquement et au débit de 12,5 Mbits/s une $m$-séquence de 255 bits, /6/. Ce signal module une porteuse de fréquence 2,2 GHz puis est filtré de manière à ce que la bande $B$ émise n'excède pas 25 MHz, et enfin amplifié. La bande $B$ analysée est centrée autour de 2,2 GHz et large de 25 MHz (au plus). La période $T$ du signal 1 vaut 20,4 $\mu s$. Le décalage en fréquence de la voie 2 vaut donc $1/2T = 24,451$ KHz. Ce décalage est ici réalisé lors de la modulation en utilisant des synthétiseurs aux fréquences respectives $f_0$ et $f_0 + 1/2T$, figure 2.

La puissance utile délivrée par chaque source est d'environ 40 dBm. Le récepteur est au mobile. Une acquisition dure $2T$=40,6 $\mu s$ et enregistre 1020 échantillons. Les points de mesure sont espacés de 2 cm et les parcours longs d'une soixantaine de mètres.

La méthode a d'abord été validée en plaçant les deux émetteurs au même endroit et en vérifiant que les réponses impulsionnelles reçues étaient effectivement identiques, figure 4. Puis d'autres mesures ont pu être conduite en situations de Macrodiversité. Elles attestent d'une excellente dynamique des réponses calculées.

Un exemple de couple de réponses impulsionnelles mesurées par ce procédé est donné en figure 5 et atteste d'une excellente dynamique des réponses calculées.

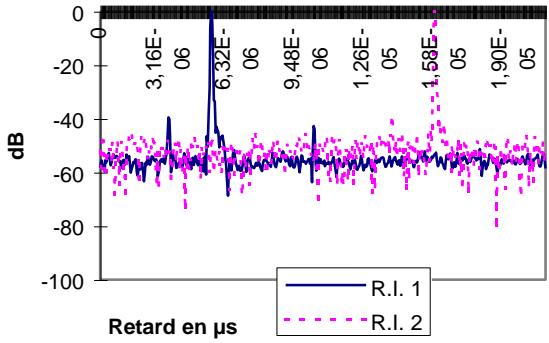

**Figure 4 : Les deux émetteurs sont confondus**

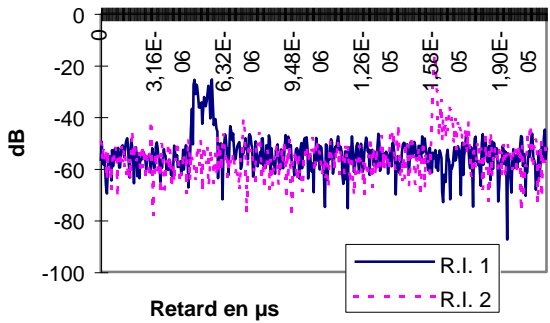

**Figure 5 : Emetteurs en Macrodiversité**

# Conclusion

Nous décrivons dans cet article une méthode novatrice de sondage pouvant être utilisée en macro ou en microdiversité. L'exposé s'accompagne d'un cas pratique et de résultats de mesure confirmant la simplicité de sa mise en œuvre et les très bonnes performances qu'on peut en attendre. Les bases théoriques de la méthode pourront être développées plus à fond et dans le cadre non restrictif d'une pluralité de canaux lors de la conférence.

Des études complémentaires ont été réalisées pour améliorer le mode opératoire et pour évaluer l'influence de certains paramètres de mesure tels le contrôle automatique de gain et la dérive temporelle ou fréquentielle des émetteurs.

La conception intégralement numérique du signal de test dédié au sondage de canal est envisagée à moyen terme, /6/. Elle reste parfaitement compatible avec la méthode proposée.

Ces différents points pourront être présentées au colloque, de même qu'une première analyse des mesures pour la modélisation du canal radiomobile en macrodiversité.

# Références

# Annexe

Plaçons-nous dans les hypothèses générales où $p$ signaux sondeurs sont émis depuis $p$ sources différentes. Intéressons-nous au canal numéro $n$, avec $1 \leq n \leq p$.

Les quotients $\dfrac{R\left(\dfrac{k}{T} + \dfrac{n-1}{pT}\right)}{S_k\left(\dfrac{k}{T} + \dfrac{n-1}{pT}\right)}$, $k \in [-N, N-1]$

fournissent une version échantillonnée en fréquences de la transformée de Fourier $H_{B,n}$ de la réponse du canal numéro $n$ vu sur la bande $B$. Nous avons par ailleurs $H_{B,n}\left(\dfrac{k}{T} + \dfrac{n-1}{pT}\right) = 0$ pour tout $k \notin [-N, N-1]$.

Effectuer une $TFD^{-1}$ de la séquence $\left(H_{B,n}\left(\dfrac{k}{T} + \dfrac{n-1}{pT}\right)\right)_{-N \leq k \leq N-1}$ revient à calculer transformée de Fourier inverse de $\sum_{k \in \mathbb{Z}} H_{B,n}\left(\dfrac{k}{T} + \dfrac{n-1}{pT}\right)\delta\left(f - \dfrac{k}{T}\right)$ puis à l'évaluer aux instants $\dfrac{iT}{2N}$.

Transformons $\sum_{k \in \mathbb{Z}} H_{B,n}\left(\dfrac{k}{T} + \dfrac{n-1}{pT}\right)\delta\left(f - \dfrac{k}{T}\right)$ en $\left(\sum_{k \in \mathbb{Z}}\delta\left(f - \dfrac{k}{T}\right)\right) \cdot \left(H_{B,n}\left(f + \dfrac{n-1}{pT}\right)\right)$, qui vaut aussi $\left(\sum_{k \in \mathbb{Z}}\delta\left(f - \dfrac{k}{T}\right)\right) \cdot \left(H_{B,n}(f) * \delta\left(f + \dfrac{n-1}{pT}\right)\right)$. Sa transformée de Fourier est donc, à une constante multiplicative près, $\left(\sum_{k \in \mathbb{Z}}\delta(t - kT)\right) * \left(h_{B,n}(t) e^{-j\frac{2\pi(n-1)}{pT}t}\right)$, soit :

$\sum_{k \in \mathbb{Z}} h_{B,n}(t - kT) e^{-j\frac{2\pi(n-1)}{pT}(t - kT)}$. D'où les résultats annoncés.